\documentclass[fleqn,10pt]{wlscirep}

\begin{document}



\title{Proposal for probing energy transfer pathway by single-molecule pump-dump experiment}

\author[1]{Ming-Jie Tao}
\author[1,*]{Qing Ai}
\author[1]{Fu-Guo Deng}
\author[2,+]{Yuan-Chung Cheng}

\affil[1]{Department of Physics, Applied Optics Beijing Area Major Laboratory,
Beijing Normal University, Beijing 100875, China}

\affil[2]{Department of Chemistry, Center for Quantum Science and Engineering,
National Taiwan University, Taipei City 106, Taiwan}

\affil[*]{aiqing@bnu.edu.cn}

\affil[+]{yuanchung@ntu.edu.tw}

\date{\today }


\begin{abstract}
The structure of Fenna-Matthews-Olson (FMO) light-harvesting complex
had long been recognized as containing seven bacteriochlorophyll
(BChl) molecules. Recently, an additional BChl molecule was
discovered in the crystal structure of the FMO complex, which may
serve as a link between baseplate and the remaining seven molecules.
Here, we investigate excitation energy transfer (EET) process by
simulating single-molecule pump-dump experiment in the
eight-molecules complex. We adopt the coherent modified Redfield
theory and non-Markovian quantum jump method to simulate EET
dynamics. This scheme provides a practical approach of detecting the
realistic EET pathway in BChl complexes with currently available
experimental technology. And it may assist optimizing design of
artificial light-harvesting devices.
\end{abstract}

\maketitle



As the chemical energy that all life on earth demand is almost from
solar energy harvested by virtue of photosynthesis, many researchers
devote themselves into improving the production of photosynthesis.
In recent decades, much attention has been focused on excitation energy transfer (EET)
in photosynthesis, that is, photosynthetic complexes
transmit efficiently the solar energy captured in the peripheral
light-harvesting antenna to reaction centres. Although the pathways
and time scales of EET are often described by semiclassical models
\cite{Blankenship,Amerongen}, we still lack precise mechanism
responsible for efficient EET. Recently, quantum coherence effects
in photosynthetic EET were predicted \cite{Knox,Leegwater} and
indirectly observed \cite{Savikhin}. In particular, as revealed by
two experiments in 2007 \cite{Engel,Lee}, the quantum coherence of
EET in natural photosynthesis has attracted more and more interest
from broad fields, such as physical society, chemical society, and
biological society. The quantum coherence manifests itself in the
wavelike evolution of exciton states \cite{Engel,Ishizaki}. In
photosynthetic systems, the quantum coherence between electronic
excitations plays an important role in the optimization of EET
efficiency \cite{Ai13}. Although much progress has been made in
revealing quantum coherence effects in the photosynthetic EET, there
is an important issue under heated debate: although it seems that
there exists the quantum coherent oscillation in site populations of
Fenna-Matthews-Olson (FMO) \cite{Harada12,Saikin14,Jia15}, others challenged this discovery as it
depended on the EET pathway and number of bacteriochlorophylls
(BChls) in FMO \cite{Ishizaki,Busch,Ritschel,Moix}.

The FMO pigment-protein complex, found in low light-adapted green
sulfur bacteria \cite{Cheng,Ishizaki}, has become an important model
system to study EET in photosynthesis
\cite{Savikhin,Engel,Ishizaki,Ai13,Cheng,Fassioli,Chin,Busch,Ritschel,Moix}.
Savikhin \emph{et al}. observed hint for quantum beating in the FMO
complex by means of pump-probe anisotropy techniques
\cite{Savikhin}. Engel \emph{et al}. studied the FMO complex
isolated from \emph{Chlorobium tepidum} with $2$D electronic
spectroscopy and gained direct evidence of long-lived electronic
coherence \cite{Engel}. Recently, the effects of quantum coherence
on enhancement of photosynthetic EET efficiency were discussed from
the perspective of quantum walk by Aspuru-Guzik and coworkers, and
by Plenio and Huelga \cite{Mohseni,Rebentrost1,Rebentrost2,Plenio},
respectively. As we know, the FMO complex is a trimer made of
identical subunits, and it had long been recognized that there are
seven BChl molecules in each monomeric subunit. However, the
presence of an additional BChl pigment in each subunit was reported
recently, which had probably been lost in the previous
recrystallization. In 2011, Busch \emph{et al}. suggested the eighth
pigment being the linker between the baseplate and the remaining
seven BChls by crystallographic studies and calculations of the
optical properties of the FMO \cite{Busch}. In the same year,
Ritschel \emph{et al}. theoretically investigated EET in the full
FMO trimer, and focused on the role of BChl 8 in the energy transfer
on different sets of transition energies. It was shown that BChl 8
plays an important role in receiving excitation from the outer light
harvesting antenna \cite{Ritschel}. Meanwhile, Moix \emph{et al}.
studied the influence of the $8$th BChl on the dynamics in FMO
through the generalized Bloch-Redfield equation and the
noninteracting blip approximation \cite{Moix}. And it was discovered
that the EET in eight-BChls complex remains efficient and robust,
although the existence of the $8$th BChl clearly affects the energy
transfer pathways as revealed by the simulated 2D electronic
spectroscopy \cite{Yeh14}. Nevertheless, these previous studies did
not come to an agreement on whether the energy flow in FMO complex
passes through site 8 or not. On the other hand, various approaches
have been put forward to optimize the energy transfer in the
eight-BChls FMO, e.g., by introducing phases in inter-site couplings
\cite{Yi13}, and by tuning temperature of the baseplate
\cite{Guan13}. At this stage, two questions naturally come to our
mind, when we consider designing artificial light-harvesting
\cite{Chen12}: whether is it necessary to include an additional BChl
in artificial light-harvesting device if the efficiency is not
essentially affected by its presence? How can we determine whether a
specific BChl is in an EET pathway or not?

Among the methods capable of detecting the ultrafast quantum
dynamics in EET, the 2D electronic spectroscopy is a
four-wave-mixing photon-echo approach to provide valuable
information about electronic transitions \cite{Engel,Cheng}.
However, as its signal is averaged over an ensemble of inhomogeneous
photosynthetic complexes, it may not be a competent candidate for
revealing the EET pathway. A recently-developed single molecule
technique has grown into a powerful method for exploring the
individual nanoscale behavior of molecules in complex local
environments \cite{Brinks2,Moerner}. In this sense,
the single-molecule pump-dump experiment provides useful
insights into the ultrafast quantum dynamics of the photosynthetic
complex \cite{Brinks,Scherer}, and thus can be used to settle this
issue of EET pathway whether site 8 is in the energy transfer
pathway of FMO. In 2005, Barbara \emph{et al}. investigated the
molecular structure and charge-transfer dynamics of conjugated
polymers through the single-molecule spectroscopy \cite{Barbara}. In
2009, Gerhardt \emph{et al}. observed 5 cycles of Rabi oscillations
in a single molecule via narrow zero-photon transition using short
laser pulses \cite{Gerhardt}. During the past decade, van Hulst and
coworkers developed single-molecule pump-probe techniques to control
vibrational wave packets and coherent energy transfer over different
pathways in individual molecules at ambient conditions
\cite{Brinks,Brinks14,Hildner13}, although it might not be easy to
distinguish single molecule coherence oscillations from spectral interference \cite{Weigel15}.
These technical advances \cite{Dijk05,Hernando06,Hildner11} inspire us to propose detecting
the EET pathway in FMO by the single-molecule pump-dump experiment.

Traditionally, the EET is described in two opposite regimes, i.e.
F\"{o}rster and Redfield theories respectively \cite{Abramavicius16}.
In order to simulate the single-molecule experiment in natural
photosynthetic complexes, we explore a theoretical approach
\cite{Ai} by combining a coherent modified Redfield theory (CMRT)
\cite{Hwang,Chang} and a non-Markovian quantum jump (NMQJ) method
\cite{Dalibard,Piilo1,Piilo2}. As a generalization of the modified
Redfield theory \cite{Zhang98,Yang02}, the CMRT can describe a
quantum system's density matrix completely \cite{Ai} and has been
successfully applied to simulate coherent EET dynamics in
photosynthetic light harvesting \cite{Novoderezhkin10,Hwang,Chang}.
In 2008, Piilo and coworkers developed an efficient
NMQJ method to simulate non-Markovian dynamics of an open quantum
system \cite{Piilo1,Piilo2}. These developments inspire us to
efficiently unravel a set of equations of motion for density matrix
by the NMQJ method \cite{Liu13,Laine14,Bylicka14}. In order to utilize
the NMQJ method, we rewrite the master equation of CMRT in the
Lindblad form \cite{Ai}. With the help of this approach, we
theoretically simulate the quantum dynamics in the single-molecule
experiment and obtain some useful results to determine whether or not
the EET in FMO passes through site 8.

\section*{Results}

\subsection*{EET-path-resolved experiment}
\label{Sec:scheme}

\begin{figure*}[htbp]             
\centering
\includegraphics[bb=30 10 580 270,width=10 cm]{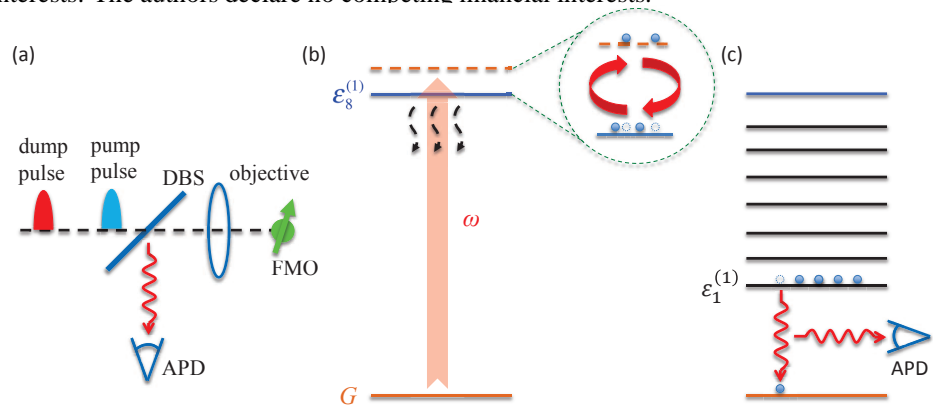}
\caption{(a) Schematic diagram of single-molecule pump-dump
experimental setup. (b) Schematic energy diagram when the dump pulse
is applied: there will be induced transition between the
highest-exciton state $\vert \varepsilon_{8}^{(1)}\rangle$ and
up-lifted ground state $\vert G\rangle$ in the rotating frame. Since
the induced-couplings between the ground state and lower-exciton
states are relatively smaller with respect to their level spacings,
they will result in small shifts to the effective energies.
Therefore, it is equivalent to a dissipative two-level system. (c)
After the pulse ends, the population of the exciton states will
relax to the lowest exciton state from which they jump to ground
state with a photon emitted.}\label{scheme}
\end{figure*}

As illustrated in Fig.~\ref{scheme}(a), we present a proposal for
settling the problem whether a site lies within the EET pathway
in a natural photosynthetic complex by single-molecule pump-dump experiment.
A mode-locked laser or optical parametric oscillator offers the visible laser pulse to induce
coherent transitions between the ground state and exciton states of
the FMO complex. The pulse train enters the confocal microscopy
which contains a dichroic beam-splitter (DBS) and an objective and
an avalanche photodiode (APD). Having passed a DBS, the pulse is
focused on a single FMO complex by an oil immersion objective. The
emitted fluorescence is collected by the same objective. After
reflected by the DBS, the fluorescence is detected by an APD. For
detailed information about the experiment setup, please refer to
Ref.~\citen{Brinks14}.

The single-molecule pump-dump (probe) technique developed by van
Hulst and coworkers has been successfully applied to resolving the
ultrafast EET dynamics at physiological
conditions~\cite{Dijk05,Hernando06}. It can effectively control
vibrational wave packets and demonstrate interference among
different EET pathways in a single photosynthetic
complex~\cite{Brinks,Hildner13}. Most importantly, a femtosecond
single-qubit operation can be carried out on single molecules at
room temperature~\cite{Hildner11}, which implies potential
application of quantum information~\cite{Ren14,Li15} on
photosynthetic light-harvesting. In this regard, the single-molecule
pump-dump technique is chosen to resolve the EET pathway in a
photosynthetic light-harvesting.

\subsubsection*{Assumptions and initial-state preparation}

Our experimental scheme is based on the following assumptions:
Without loss of generality, the initial excitation of FMO is
prepared at site 1 or site 8 for the sake of simplicity in our
numerical simulation. In order to realize this assumption, the total
system is made up of an FMO, a baseplate, and an antenna. The
initial excitation is prepared at the antenna by absorbing a photon
at the frequency with a blue shift to the highest exciton state of
FMO. Furthermore, the antenna is spatially far away from the FMO in
order not to excite the FMO during the preparation process when the
pump pulse is applied. Since the total Hamiltonian including the FMO
and the baseplate and outer antenna is unknown, the initial
excitation at site 1 or site 8 in FMO is utilized to simplify
numerical simulation.
Moreover, we further assume that the site energy for the eighth site
of FMO is the highest of all sites. This seems to be true according
to simulations made in Ref.~\citen{Busch}. Due to its small
electronic couplings to other sites, the highest exciton state is a
localized eigen state of FMO Hamiltonian. In this case, the energy
flow through site 1 only will not essentially pass through site 8.

\subsubsection*{Experimental scheme}

As shown in Fig.~\ref{scheme}(a), based on the above assumptions, we
propose the following experimental scheme to determine the EET path
in FMO, including the main procedures as follows:

After photoexcitation by the pump pulse, a single-molecule FMO
evolves freely from the initial site, i.e., site 1 or site 8. Then,
a dump pulse begins to be applied to the FMO molecule at time $t_1$.
Its driving frequency $\omega$ is chosen to be in close resonance
with the transition from the ground state to the highest exciton
state, meanwhile it is largely detuned from the transitions to other
exciton states. In this case, the population on the target state can
be coherently transferred to the ground state, while the population
on the other exciton states can be nearly undisturbed. The dump
pulse ends at time $t_2$ and then the FMO molecule is left to evolve
freely again. Due to the coupling to the phonon bath, the remaining
population on the single-excitation subspace quickly relaxes to the
lowest exciton state without emitting a photon. Finally, the
population on the lowest exciton state transits to the ground state
with fluorescence detected by photon detector. Generally speaking,
the relaxation within the single-excitation subspace, e.g.
$0.1\sim10$ ps, is much faster than dissipation to the ground state,
e.g. $\sim1$ ns. Therefore, we would expect that all the population
on the higher exciton states could reach the lowest exciton state
before they emit a fluorescent photon.

\subsubsection*{Parameter optimization}

Since we aim at transferring the population on the highest exciton
state $\vert\varepsilon_{8}^{(1)}\rangle$ to the ground state
$\vert\varepsilon_{0}^{(1)}\rangle=\vert G\rangle$, we should tune
the driving frequency $\omega$ in close resonance with the target
state $\vert\varepsilon_{8}^{(1)}\rangle$ while let it be largely
detuned from other exciton states $\vert
\varepsilon_{k}^{(1)}\rangle (k=1,\ldots7)$. In order to fulfill
this requirement, the driving frequency is chosen as the one with a
blue shift to the transition between
$\vert\varepsilon_{8}^{(1)}\rangle$ and
$\vert\varepsilon_{0}^{(1)}\rangle$, i.e.,
$\omega\geq\varepsilon_{8}^{(1)}-\varepsilon_{0}^{(1)}=12709~\rm{cm^{-1}}$.

Furthermore, the electric polarization of the dump pulse is chosen
at the direction perpendicular to both $\hat{r}_{60}$ and
$\hat{r}_{70}$ where $\hat{r}_{k0}$ is the unit vector of the
transition dipole between the $k$th exciton state and the ground
state~\cite{Moix}, i.e., $\vec{E}\parallel(0.680\textrm{,}0.323\textrm{,}0.658)$.
In this case, there will induce the couplings between the
ground state and the exciton states with strengths
\begin{eqnarray}
g_{k0}=q\vec{r}_{k0}\cdot\vec{E}
=qE\!\times\! (5.52\textrm{,}10.3\textrm{,}-6.94\textrm{,}-5.53\textrm{,}-2.26\textrm{,}0\textrm{,}0\textrm{,}10)\textrm{au}.\;\;\;
\end{eqnarray}

The transition dipole orientations $\hat{r}_{k0}$ can be measured by
the reduced linear dichroism signal, which is calculated from
 two orthogonal polarizations detected in the in-plane of the sample
\cite{Fourkas01}. This technique was theoretically proposed
by Fourkas \cite{Fourkas01}, and experimentally realized by Vacha
\cite{Vacha03}. On account of the large-detuning condition, i.e.,
$\omega+\varepsilon_{0}^{(1)}-\varepsilon_{k}^{(1)}\geq10g_{k0}$ for
$k=1,\ldots7$, the maximum coupling between the highest exciton
state and the ground state is $g_{80}=35~\rm{cm^{-1}}$. Notice that
a strong Rabi frequency as large as 318 cm$^{-1}$ was realized in
experiments \cite{Brinks14}. Since the typical transition dipoles of
BChls are of the order of several Debyes, the maximum coupling
induced by laser fields is achievable in practice. For numerical
simulations, the information for all molecular electric dipoles
$\vec{r}(n)$ are provided in the Appendix.

\subsection*{Theoretical simulation method}
\label{Method}

We adopt a Frenkel exciton model to describe photoexcitations in the
FMO complex \cite{Cheng}. The model includes electronic interactions
between any two sites of FMO, and the system Hamiltonian is~\cite{Moix}
\begin{eqnarray}
H_{S}^{(1)}=\sum_{n=1}^{8} [E_{n}\left\vert
n\right\rangle \left\langle n\right\vert +\sum_{m=1(\neq
n)}^{8}J_{mn}\left\vert m\right\rangle \left\langle n\right\vert
 ] +E_{0}\left\vert G\right\rangle \left\langle G\right\vert ,
\end{eqnarray}
where $\vert n\rangle$ is the state with single-excitation on the
$n$th site, $E_{n}$ is the corresponding site energy, and $J_{mn}$
is the electronic coupling between site $m$ and $n$. Besides, there
is no excitation on the ground state $\vert G\rangle$ with energy
$E_{0}$ and there is no direct coupling between it and the
single-excitation states. We adopt the effective Hamiltonian for the
8-sites FMO proposed by Moix \textit{et al.}, which provides
excellent description of the spectra and EET dynamics of the
complex. The explicit form of Hamiltonian $H_{S}^{(1)}$ is given in
the Appendix.

To describe the EET dynamics induced by the system-bath couplings,
we could obtain the master equation for the CMRT as
\cite{Hwang,Chang}
\begin{equation}
\partial_{t}\rho^{(1)}  \!\!=\!\! -i [H_{e}^{(1)},\rho^{(1)} ]  -\frac{1}{2}\sum_{k\neq k^{\prime}}R_{kk^{\prime}}^{(1)\textrm{dis}}(t)
 [ \{A_{kk^{\prime}}^{(1)\dagger}A_{kk^{\prime}}^{(1)},\rho^{(1)} \}
-2A_{kk^{\prime}}^{(1)}\rho^{(1)}A_{kk^{\prime}}^{(1)\dagger} ]      -\sum_{k\neq k^{\prime}}R_{kk^{\prime}}^{(1)\textrm{pd}}(t)
        \rho_{kk^{\prime}}^{(1)} \vert \varepsilon_{k}^{(1)} \rangle  \langle \varepsilon_{k^{\prime}}^{(1)} \vert .\label{eq3}
\end{equation}
Here $H_{e}^{(1)}$ governs the coherent evolution of the EET. It
owns the same eigen states $\vert \varepsilon_{k}^{(1)}\rangle$ as
$H_{S}^{(1)}$ but with different eigen energies
$\varepsilon_{k}^{(1)}=\varepsilon_{k}^\prime-\sum_{n=1}^{8} [a_{kk}^{(1)}(n) ]^2\lambda_n$,
where the eigen energies $\varepsilon_{k}^\prime$ of $H_{S}^{(1)}$
are modified by the reorganization energies induced by the
system-bath couplings, and
$a_{kk^\prime}^{(1)}(n)=C^{\ast}_k(n)C_{k^\prime}(n)$ is the overlap
of $k$th and $k^\prime$th eigen states at site $n$. Notably, the
equation of motion is in a generalized Lindblad form \cite{Ai} with
the jump operators defined as
$A_{kk^{\prime}}^{(1)}\;\;=\;\; \vert
\varepsilon_{k}^{(1)} \rangle  \langle
\varepsilon_{k^{\prime}}^{(1)} \vert$.
Based on the CMRT, the dissipation and pure-dephasing rates are respectively
\begin{eqnarray}
R_{kk^{\prime}}^{(1)\textrm{dis}}\!\!\!&=&\!\!\!2\textrm{Re}\int_{0}^{t}d\tau
e^{i(\varepsilon_{k^{\prime}}^{(1)}
-\varepsilon_{k}^{(1)})t}e^{-[g_{kkkk}^{(1)}(\tau)
+g_{k^{\prime}k^{\prime}k^{\prime}k^{\prime}}^{(1)}(\tau)-2g_{kk,k^{\prime}k^{\prime}}^{(1)}(\tau)]
-i(\lambda_{kkkk}^{(1)}+\lambda_{k^{\prime}k^{\prime}k^{\prime}k^{\prime}}^{(1)}-2\lambda_{kkk^{\prime}k^{\prime}}^{(1)})\tau}      \nonumber   \\
\;\;&& \times\{\ddot{g}_{k^{\prime}kkk^{\prime}}^{(1)}(\tau)
-[\dot{g}_{k^{\prime}kkk}^{(1)}(\tau)-\dot{g}_{k^{\prime}kk^{\prime}k^{\prime}}^{(1)}(\tau)-2i\lambda_{k^{\prime}kk^{\prime}k^{\prime}}^{(1)}]    \times[\dot{g}_{kk^{\prime}kk}^{(1)}(\tau)-\dot{g}_{kk^{\prime}k^{\prime}k^{\prime}}^{(1)}(\tau)-2i\lambda_{kk^{\prime}k^{\prime}k^{\prime}}^{(1)}]\},   \nonumber 
\label{eq5}
\end{eqnarray}
\begin{eqnarray}
R_{kk^{\prime}}^{(1)\textrm{pd}}(t) \!\!&=&\!\! \sum_{n=1}^{8} [a_{kk}^{(1)}(n)
-a_{k^{\prime}k^{\prime}}^{(1)}(n) ]^{2}\textrm{Re} [\dot{g}_{n}(t) ],\label{eq6}
\end{eqnarray}
where
\begin{eqnarray}
g_{k_{1}k_{2}k_{3}k_{4}}^{(1)}(t)   = \sum_{n=1}^{8}a_{k_{1}k_{2}}^{(1)}(n)a_{k_{3}k_{4}}^{(1)}(n)g_{n}(t),
\lambda_{k_{1}k_{2}k_{3}k_{4}}^{(1)} = \sum_{n=1}^{8}a_{k_{1}k_{2}}^{(1)}(n)a_{k_{3}k_{4}}^{(1)}(n)\lambda_{n},
\end{eqnarray}
\begin{eqnarray}
g_{n}(t)    \!\!&=&\!\! \int d\omega\frac{J_{n}(\omega)}{\omega^{2}} [ (1-\cos\omega t )\coth (\frac{\beta\omega}{2} )
+i (\sin\omega t-\omega t ) ]
\end{eqnarray}
is the lineshape function. $\lambda_{n}$ and $J_n(\omega)$ are the reorganization energy and
spectral density of the $n$th molecule, respectively. $\beta=1/k_BT$
with $k_B$ and $T$ are Boltzman constant and temperature,
respectively. In our numerical simulations, we assume identical
reorganization energy $\lambda=35$ cm$^{-1}$ and identical Ohmic
spectral density
$J(\omega)=\lambda(\omega/\omega_c)\exp(-\omega/\omega_c)$ with
cut-off $\omega_c=50$ cm$^{-1}$ for all molecules and the experiment
is conducted at ambient temperature, i.e. $T=300$ K.

In order to implement the NMQJ method, we rewrite Eq.~(\ref{eq3}) in
the Lindblad form as \cite{Ai}
\begin{equation}
\partial_{t}\rho^{(1)} \;\; =\;\;   -i [H_{e}^{(1)},\rho^{(1)} ]
-\frac{1}{2}\sum_{k,k^{\prime}}R_{kk^{\prime}}^{(1)}(t) [ \{ A_{kk^{\prime}}^{(1)\dagger}A_{kk^{\prime}}^{(1)},\rho^{(1)} \}
-2A_{kk^{\prime}}^{(1)}\rho^{(1)}A_{kk^{\prime}}^{(1)\dagger} ],
\end{equation}
where the matrix element of the rates is defined as
\begin{equation}
R_{kk^{\prime}}^{(1)}  \;\; \equiv\;\;  \begin{cases}
\begin{array}{cc}
R_{kk^{\prime}}^{(1)\textrm{dis}}, & k\neq k^{\prime}   \\
\Gamma_{k}^{(1)}, & k=k^{\prime}
\end{array}.
\end{cases}
\end{equation}
The dephasing rates are given by
\begin{equation}
\Gamma^{(1)}\;\;=\;\;M^{-1}B^{(1)},
\end{equation}
where the matrix elements of $B^{(1)}$ and $M$ are respectively
\begin{equation}
B_{a}^{(1)}\;\;=\;\;\sum_{k=a+1}^{8}R_{ak}^{(1)\textrm{pd}}+\sum_{k=1}^{7}R_{ka}^{(1)\textrm{pd}},
M_{jk} \;\; =\;\;   \begin{cases}
\begin{array}{cc}
\frac{1}{2}, & k<j              \\
\frac{1}{2}(16-j), & k=j        \\
1, & j<k<8                      \\
\frac{1}{2}, & \textrm{otherwise}
\end{array}.
\end{cases}
\end{equation}

As shown in Fig.~\ref{scheme}(b), after the free evolution, there is
a dump pulse with frequency $\omega$ applied to the FMO molecule and
thus transitions between the ground state $\vert
\varepsilon_{0}^{(1)}\rangle$ and delocalized exciton states $\vert
\varepsilon_{k}^{(1)}\rangle(k\neq0)$ are induced. In this
situation, the electronic Hamiltonian reads
\begin{eqnarray}
H_{S}^{(2)}=\sum_{k=0}^{8}\varepsilon_{k}^{\prime} \vert
\varepsilon_{k}^{(1)} \rangle  \langle
\varepsilon_{k}^{(1)} \vert +2\cos\omega
t\sum_{k=1}^{8}g_{k0} ( \vert
\varepsilon_{k}^{(1)} \rangle  \langle
\varepsilon_{0}^{(1)} \vert +\textrm{h.c.} ),
\end{eqnarray}
where $2g_{k0}$ is the laser-induced coupling strength between the
ground state $\vert \varepsilon_{0}^{(1)}\rangle$ and the
delocalized exciton state $\vert \varepsilon_{k}^{(1)}\rangle$.

Transformed to a rotating frame with $U=\exp[i\omega t\vert
\varepsilon_{0}^{(1)}\rangle \langle \varepsilon_{0}^{(1)}\vert ]$,
the effective Hamiltonian of the electronic part
$H_{\textrm{eff}}^{(2)}    =
U^{\dagger}H_{S}^{(2)}U+i\dot{U}^{\dagger}U$ is
\begin{eqnarray}
H_{\textrm{eff}}^{(2)}   \simeq
\sum_{k=1}^{8}\varepsilon_{k}^{\prime} \vert
\varepsilon_{k}^{(1)} \rangle  \langle
\varepsilon_{k}^{(1)} \vert
+ (\varepsilon_{0}^{\prime}+\omega ) \vert
\varepsilon_{0}^{(1)} \rangle  \langle
\varepsilon_{0}^{(1)} \vert
+\sum_{k=1}^{8}g_{k0} ( \vert
\varepsilon_{k}^{(1)} \rangle  \langle
\varepsilon_{0}^{(1)} \vert +\textrm{h.c.} ),\;\;\;\;\;\;
\end{eqnarray}
where we have dropped the fast-oscillating terms with factors $\exp\left(\pm i2\omega t\right)$. The above Hamiltonian can be diagonalized as
$H_{\textrm{eff}}^{(2)}=\sum_{k=0}^{8}\varepsilon_{k}^{\prime\prime}\vert \varepsilon_{k}^{(2)}\rangle \langle \varepsilon_{k}^{(2)}\vert$,
where
$\vert \varepsilon_{k}^{(2)}\rangle =\sum_{k^{\prime}=0}^{8}C_{kk^{\prime}}^{(2)}\vert \varepsilon_{k^{\prime}}^{(1)}\rangle
=\sum_{n=1}^{8}\sum_{k^{\prime}=0}^{8}C_{kk^{\prime}}^{(2)}C_{k^{\prime}n}^{(1)}\vert n\rangle$
is the eigen state with eigen energy $\varepsilon_{k}^{\prime\prime}$.

In the basis of $\{ \vert \varepsilon_{k}^{(2)}\rangle,k=0,1,\cdots8
\}$ of the rotating frame, we could obtain the master equation of
the same form as Eq.~(\ref{eq3}), but the system Hamiltonian is
replaced by
$H_{e}^{(2)}=\sum_{k=0}^{8}\varepsilon_{k}^{(2)}\vert\varepsilon_{k}^{(2)}\rangle\langle\varepsilon_{k}^{(2)}\vert$
with eigen energies
$\varepsilon_{k}^{(2)}=\varepsilon_{k}^{\prime\prime}-\sum_{n=0}^{8} [a_{kk}^{(2)}(n) ]^2\lambda_n$.
The dissipation and pure-dephasing rates can be calculated in the same way as Eq.~(\ref{eq5}) but $a_{kk^{\prime}}^{(1)}(n)$ is substituted by
$a_{kk^{\prime}}^{(2)}(n)    =   \sum_{k_{1}=0}^{8}C_{kk_{1}}^{(2)*}C_{k_{1}n}^{(1)*}\sum_{k_{2}=0}^{8}C_{k^{\prime}k_{2}}^{(2)}C_{k_{2}n}^{(1)}$.
As a consequence, for the duration with a pulse, the master equation
can also be rewritten in the Lindblad form and thus be solved by the
NMQJ approach. It is worthy of mentioning that the calculated
density matrix should be transformed back to the static frame as
$\rho=U\rho^{(2)}U^{\dagger}$. Note that in order to test
experimentally the EET pathway through site 8 in the FMO, here we
assume the laser frequency is in resonance with
$\vert\varepsilon_0^{(1)}\rangle\leftrightarrow\vert\varepsilon_8^{(1)}\rangle$.
However, our formulism is general and could be applied to other
resonance conditions.

\subsection*{Results and analysis}
\label{Results}

In the previous section, we briefly introduced a newly-developed
CMRT-NMQJ approach \cite{Ai,Hwang} to simulate the quantum dynamics
of an FMO complex in a single-molecule pump-dump experiment, as
presented in Fig.~\ref{scheme}.

\subsubsection*{Exact result}

Since a portion of population of the exciton states has been
transferred to the ground state during the dump pulse duration, the
detected fluorescence intensity is determined by the population on
the ground state, which can be controlled by the following
parameters, i.e., the Rabi frequency $g\equiv g_{80}$, the beginning
time of the laser pulse $t_{1}$, the pulse width $T=t_2-t_1$, and
the frequency of drive $\omega$. Before the numerical simulation, we
define the fluorescence quantum yield as
\begin{equation}
\Phi (t_{1},T,\omega,g )=1- \langle G \vert\rho^{(1)}(t_{1}+T) \vert G \rangle,
\end{equation}
which is the total population on the single-excitation subspace
right after the dump pulse ends. As the laser field is tuned in
close resonance with the selected energy level, the fluorescence
quantum yield is very sensitive to the irradiation condition. In
order to clearly illustrate this phenomenon and also compare the
effects of different initial states, for a given range of
experimental parameters, we can define the visibility of
fluorescence quantum yield as
\begin{equation}
V=\frac{\max (\Phi )-\min (\Phi )}{\max (\Phi )+\min (\Phi )}.
\end{equation}

\begin{figure}
\centering
\includegraphics[width=11 cm]{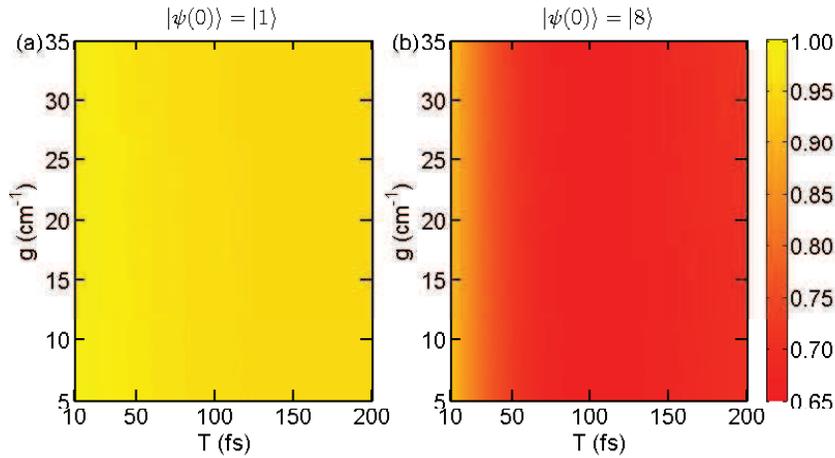}
\caption{Quantum yield $\Phi$ vs the pulse duration $T$ and Rabi
frequency $g$ for the driving frequency
$\omega=12709~\textrm{cm}^{-1}$ and the beginning time
$t_{1}=50~\textrm{fs}$: (a) initial state $\vert \psi(0)\rangle
=\vert 1\rangle$ with observed visibility $V=0.02$; (b) $\vert
\psi(0)\rangle =\vert 8\rangle$ with $V=0.16$.}\label{GT}
\end{figure}

\begin{figure}
\centering
\includegraphics[width=11 cm]{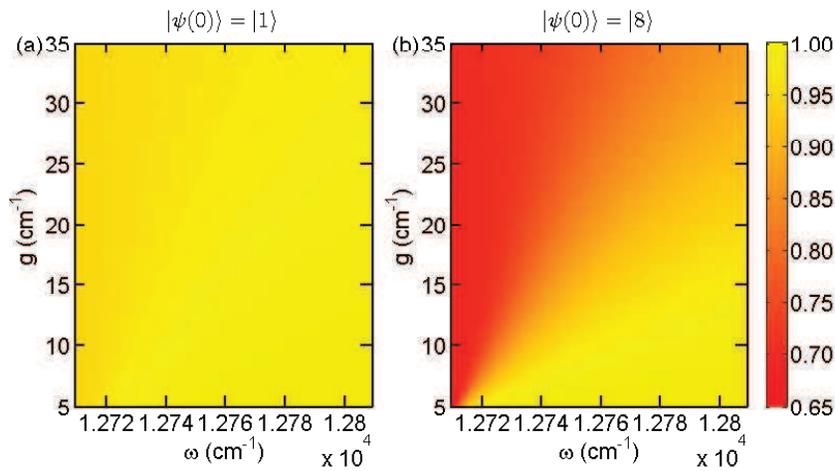}
\caption{Quantum yield $\Phi$ vs the driving frequency
$\omega$ and Rabi frequency $g$ for the pulse duration
$T=170~\textrm{fs}$ and the beginning time
$t_{1}=50~\textrm{fs}$: (a) initial state $\vert \psi(0)\rangle
=\vert 1\rangle$ with fluorescence visibility $V=0.03$; (b) $\vert
\psi(0)\rangle =\vert 8\rangle$ with $V=0.18$.}\label{GOmega}
\end{figure}

As shown in Fig.~\ref{GT}, we plot the quantum yield $\Phi$ vs the
pulse width $T$ and Rabi frequency $g$ for the driving frequency
$\omega=12709~\rm{cm^{-1}}$ and the beginning time $t_{1}=50$ fs. In
Fig.~\ref{GT}(a), for the whole parameter range, the quantum yields
are very close to unity, since the applied laser field is largely
detuned from the state $\vert 1\rangle\simeq0.55\vert
\varepsilon_3^{(1)}\rangle+0.78\vert\varepsilon_6^{(1)}\rangle$. In
this case, the population on lower exciton states cannot be
transferred to the ground state by means of the laser. In contrast,
as illustrated in Fig.~\ref{GT}(b), the detected quantum yields vary
significantly for different Rabi frequencies $g$ and pulse width
$T$. Especially, for a given Rabi frequency, the quantum yield
decreases along with the increase of pulse width. That is because
more population on the highest exciton state will be resonantly
transferred to the ground state through the laser-induced transition
as the pulse lasts for a longer duration. Obviously, there is a
larger visibility for the case with the initial state
$\vert\psi(0)\rangle
=\vert8\rangle\simeq\vert\varepsilon_8^{(1)}\rangle$ in comparison
to that with $\vert\psi(0)\rangle=\vert1\rangle$, i.e., $V=0.16$ vs
$V=0.02$, due to the close-resonance condition. This remarkable
difference confirms our conjecture that the difference in initial
states can be detected by the single-molecule pump-dump experiment.
A similar result, i.e., $V=0.18$ vs $V=0.03$, is also observed for
the quantum yield $\Phi$ vs the driving frequency $\omega$ and the
Rabi frequency $g$ for the pulse duration $T=170$ fs and the
beginning time $t_{1}=50$ fs in Fig.~\ref{GOmega}. We also notice
that the dependence of $\Phi$ on $\omega$ and $g$ becomes more
complicated, shown in Fig.~\ref{GOmega}(b). When we increase the
drive frequency, the quantum yield of fluorescence also raises as
the laser field is tuned far away from resonance with the highest
exciton state. In this situation, no population on the exciton
states can be effectively transferred to the ground state. Besides,
for a realistic molecule, there would be fluctuations in the site
energies, or the laser field is not exactly perpendicular to the
selected transition dipoles, i.e., $g_{60},g_{70}\neq0$. Even in
this case, according to our numerical simulation, the expected
visibility for the initial state $\vert8\rangle$ is still
significantly larger than that for $\vert 1\rangle$. As a practical
criterion, we set the median $V_{m}\simeq0.1$ to judge whether the
EET path is through $\vert8\rangle$ or not. For a visibility larger
than $V_{m}$, the energy is transferred through $\vert8\rangle$,
otherwise it is through $\vert1\rangle$ only.

\subsubsection*{Dissipative two-level system approximation}

In order to reveal the underlying physical mechanism, we
approximately describe the above experiment by a dissipative
two-level system in the closely-resonant case. To be specific, the
laser is tuned in close resonance with the transition between the
highest-exciton state and the ground state, i.e.,
$g_{80}\gtrsim\vert\omega+\varepsilon_0^{(1)}-\varepsilon_8^{(1)}\vert$.
In this case, the system is governed by two sets of differential
equations. For the duration of free evolution, that is
\begin{eqnarray}
\partial_{t}\rho_{00}^{(1)} = 0,
\partial_{t}\rho_{88}^{(1)} \simeq-\sum_{k=1}^{7}R_{k8}^{(1)}\rho_{88}^{(1)},
\partial_{t}\rho_{80}^{(1)} =0.
\end{eqnarray}
Straightforwardly, at the end of the free evolution, we have
\begin{eqnarray}
\rho_{00}^{(1)}(t_{1})  =   \rho_{80}^{(1)}(t_{1})=0,    \rho_{88}^{(1)}(t_{1})
=   \rho_{88}^{(1)}(0)\exp [-\int_{0}^{t_1}\sum_{k=1}^{7}R_{k8}^{(1)}(t)dt ].
\end{eqnarray}
In this situation, the population of the ground state remains
unchanged because there is no transition to the ground state induced
by either the laser pulse or the system-bath couplings. And the loss
of population in the highest exciton state results from its
dissipation to the lower exciton states.

On the other hand, for the duration with a laser pulse applied, the
equation of motion for the two-level system reads
\begin{eqnarray}
\partial_{t}\rho_{00}^{(2)}
\simeq-\sum_{k=1}^{8}R_{k0}^{(2)}\rho_{00}^{(2)}+R_{08}^{(2)}\rho_{88}^{(2)},
\partial_{t}\rho_{88}^{(2)}
\simeq-\sum_{k=0}^{7}R_{k8}^{(2)}
\rho_{88}^{(2)}+R_{80}^{(2)}\rho_{00}^{(2)},
\partial_{t}\rho_{80}^{(2)} = - [\frac{1}{2}\sum_{k=0}^{8} (R_{k0}^{(2)}
+R_{k8}^{(2)} )+i (\varepsilon_{8}^{(2)}-\varepsilon_{0}^{(2)}
 ) ]\rho_{80}^{(2)}. \label{eq30}
\end{eqnarray}
On account of the unitary transformation from the eigen bases in the
static frame $\{\varepsilon_k^{(1)}\}$ to the eigen bases in the
rotating frame $\{\varepsilon_k^{(2)}\}$, i.e.,
$\rho^{(2)}=Q^{\dagger}U^{\dagger}\rho^{(1)}UQ$,
with
\begin{equation}
Q=\left(\begin{array}{cc}
\cos\frac{\alpha}{2} & -\sin\frac{\alpha}{2}    \\
\sin\frac{\alpha}{2} & \cos\frac{\alpha}{2}
\end{array}\right)
\end{equation}
in the bases $\{ \vert \varepsilon_{8}^{(1)}\rangle ,\vert \varepsilon_{0}^{(1)}\rangle \}$, the initial condition is
\begin{eqnarray}
\rho_{88}^{(2)}(t_{1})  =
\rho_{88}^{(1)}(t_{1})\cos^{2}\frac{\alpha}{2},\;\;\;\;
\rho_{00}^{(2)}(t_{1})  =
\rho_{88}^{(1)}(t_{1})\sin^{2}\frac{\alpha}{2},\;\;\;\;
\rho_{80}^{(2)}(t_{1})  =
\rho_{88}^{(1)}(t_{1})\frac{\sin\alpha}{2}. \notag
\end{eqnarray}
Here, the mixing angle is defined as
$\tan\alpha=2g_{80}/(\varepsilon_{8}^{\prime}-\varepsilon_{0}^{\prime}-\omega)$.
At the end of the pulse duration, the coherence between $\vert\varepsilon_{0}^{(2)}\rangle$ and $\vert\varepsilon_{8}^{(2)}\rangle$ is given by
\begin{equation}
\rho_{80}^{(2)}(t_{2})=\rho_{80}^{(2)}(t_{1})\exp \{-\int_{t_{1}}^{t_{2}} [\frac{1}{2}\sum_{k=0}^{8} (R_{k0}^{(2)}(t)
+R_{k8}^{(2)}(t) )+i (\varepsilon_{8}^{(2)}
-\varepsilon_{0}^{(2)} ) ]dt \},
\end{equation}
and $\rho_{88}^{(2)}(t_2)$ and $\rho_{00}^{(2)}(t_2)$ are obtained by solving the first two equations of Eq.~(\ref{eq30}).

\begin{figure}
\centering
\includegraphics[width=11 cm]{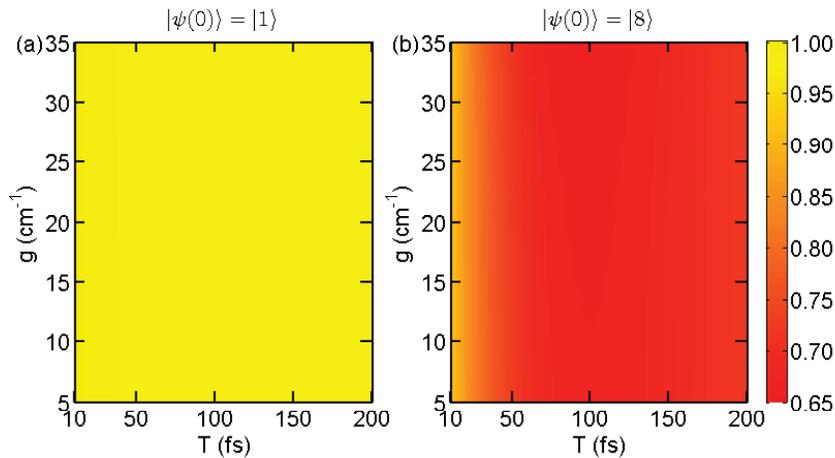}
\caption{Quantum yield $\Phi$ vs $T$ and $g$ for the dissipative
two-level system approximation: (a) initial state $\vert
\psi(0)\rangle =\vert 1\rangle$ with $V=0.006$; (b) $\vert
\psi(0)\rangle =\vert 8\rangle$ with $V=0.16$. All the parameters
are the same as those used in Fig.~\ref{GT}.}\label{GTapp}
\end{figure}

Based on the dissipative two-level approximation, the numerical
evaluation of quantum yield $\Phi$ vs Rabi frequency $g$ and pulse
width $T$ is given in Fig.~\ref{GTapp}. At the first glance, the two
sub-figures are similar to the counterparts in Fig.~\ref{GT}. As
shown in Fig.~\ref{GTapp}(b), when we gradually increase the width
from zero, the quantum yield quickly drops to nearly one half of the
original value. Moveover, as the longer the dump pulse lasts, the
more population on $\vert\varepsilon_8^{(1)}\rangle$ will dumped to
the ground state, leading to lower fluorescence yield. However,
because the large-detuning condition sets a upper bound for the Rabi
frequency, this effectively prevents complete population transfer to
the ground state. Moreover, if we further raise the Rabi frequency,
the discrepancy between Fig.~\ref{GTapp}(b) and Fig.~\ref{GT}(b)
will become significant, which is not shown here, since the
two-level approximation breaks down in this situation and more
exciton states will be probed by the laser. On the other hand, if we
extend the pulse duration further, the quantum yield will rise again
due to Rabi oscillation. However, dissipation to lower exciton
states will eventually erase the Rabi oscillation. This discovery is
consistent with our conjecture that the FMO complex prepared
initially at $\vert8\rangle$ experiences Rabi oscillation between
the highest exciton state and the ground state when there is a laser
pulse applied on the system.

\section*{Discussion}
\label{Discussion}

As demonstrated in Ref.~\citen{Mitra}, by reducing the number of
chlorophylls, the photosynthesis can be optimized in design of
artificial light-harvesting. When a chlorophyll is not within the
EET pathway, it is redundant to the light-harvesting device. Here,
we propose a single-molecule pump-dump experiment scheme for
detecting the EET pathway in BChl complexes. By coherently dumping
the population on the exciton-states in the EET pathway to the
ground state, the energy transfer path in FMO can be determined by
detecting fluorescence emission. For a smaller fluorescence
visibility $V$, it corresponds to the EET path through site $1$
only, as the energy flow through this path has not been probed by
the laser due to the large-detuning condition. On the contrary, for
a larger $V$, the EET passes through the site $8$, because the
amount of fluorescence can be tuned by adjusting the frequency and
width of the pulse, and the Rabi frequency induced by the laser.

In order to simulate the quantum dynamics in the single-molecule
pump-dump experiments, we utilize the newly-developed CMRT-NMQJ
approach \cite{Ai,Ai13}. The CMRT describes the quantum dynamics of
EET in photosynthetic complexes over a broad parameter regime
\cite{Hwang,Chang} and it is generalized to simulate the energy
transfer in the presence of laser fields \cite{Ai}. Furthermore, the
master equation of CMRT recast in Lindblad form can be efficiency
solved by the NMQJ method \cite{Piilo1,Piilo2}. Since the CMRT can
simulate the absorption spectrum, it can self-consistently obtain
the parameters for further simulation \cite{Ai}. In a recent paper \cite{Kimura15},
due to quantum vibrational effects the transfer rate is smaller than
that obtained from modified Redfield theory, which may imply that
it will take more time for the energy transfer from BChl 8. In this sense,
the difference between the visibilities in the cases with or without BChl 8
would probably become more notable. On the other hand, in Ref.~\citen{Hwang},
the population dynamics for FMO are compared by the CMRT and HEOM. Clearly,
the coherent dynamics simulated by the HEOM can be well reproduced by the CMRT.
Particularly, the population transfer time therein is consistent with that
obtained by the HEOM. However, the observable discrepancy lies in the
steady-state population. Together with a recently-developed improved variational
master equation theory \cite{Fujihashi14}, this problem could be well fixed.

We further remark that the laser-induced Rabi frequency $g_{80}$
between the highest exciton state $\vert
\varepsilon_{8}^{(1)}\rangle$ and the ground state $\vert
\varepsilon_{0}^{(1)}\rangle$ should be sufficiently large when it
is compared to their effective level spacing
$|\varepsilon_{8}^{(1)}-\varepsilon_{0}^{(1)}-\omega|$   for a
sufficient amount of population to be transferred to the ground
state. Meanwhile, the laser-induced Rabi frequencies $g_{k0}$
between other exciton states $\vert \varepsilon_{k}^{(1)}\rangle$
($k=1,2\cdots7$) and the ground state
$\vert\varepsilon_{0}^{(1)}\rangle$ should be sufficiently small
when compared to their effective level spacings
$|\varepsilon_{k}^{(1)}-\varepsilon_{0}^{(1)}-\omega|$ in order to
 not transfer the population on the other exciton states to the ground
state. In this case, the FMO complex under the quantum control of
laser fields experiences Rabi oscillation and relaxation due to the
couplings to the bath. In other words, the complex can be well
described by a dissipative two-level system under the influence of
laser pulses. On the other hand, our scheme is based on the sample
where an FMO complex and a baseplate and outer antenna get together.
Since it might not be easy to fabricate such a compound complex, an
alternative way is to prepare the single-excitation states in the
site basis, i.e., $\vert1\rangle$ and $\vert8\rangle$. By using a
combination of laser pulses with different physical parameters,
e.g., frequency, width, and amplitude, we can effectively prepare
the FMO complex in such states with a specific site excited.
However, since it is beyond the scope of the current paper, the
scheme for state preparation will be presented in a forthcoming
paper.

\section*{Methods}

The $n$th electric dipole
$q\vec{r}(n)=q [\vec{r}_{N_d}(n)-\vec{r}_{N_b}(n) ]$ points
along the axis connecting the $N_b$ and $N_d$ atoms of $n$th BChl
molecule \cite{Moix}, where
\begin{equation}
\vec{r}_{N_{d}}=\left(\begin{array}{rrr}
53.115 & 57.663 & 22.495\\
57.544 & 53.598 & 32.966\\
51.410 & 44.665 & 44.744\\
38.826 & 41.519 & 44.883\\
35.620 & 46.497 & 31.179\\
39.775 & 47.059 & 23.300\\
47.164 & 44.038 & 35.135\\
36.872 & 28.084 & 14.773
\end{array}\right){\AA},\;\;\;\;\;\; \vec{r}_{N_{b}}=\left(\begin{array}{rrr}
53.010 & 58.800 & 18.681\\
54.512 & 56.023 & 31.875\\
47.671 & 44.909 & 46.141\\
38.823 & 43.087 & 41.219\\
32.683 & 49.171 & 31.372\\
43.176 & 48.530 & 21.901\\
47.865 & 43.871 & 31.216\\
32.934 & 27.083 & 14.955
\end{array}\right){\AA}
\end{equation}
can be obtained from Ref.~\citen{Tronrud09}. As a result, the transition dipole between the ground state and $k$th exciton state reads
$q\vec{r}_{k0}=q\sum_{n=1}^8 C_k(n)\vec{r}(n)$.

According to Ref.~\citen{Moix}, the Hamiltonian for eight-BChls FMO in the single-excitation subspace is
\begin{equation}
H_{\textrm{SES}}=\left(\begin{array}{rrrrrrrr}
 310.0 & -97.9 &   5.5 &  -5.8 &   6.7 & -12.1 & -10.3 &  37.5  \\
 -97.9 & 230.0 &  30.1 &   7.3 &   2.0 &  11.5 &   4.8 &   7.9  \\
   5.5 &  30.1 &     0 & -58.8 &  -1.5 &  -9.6 &   4.7 &   1.5  \\
  -5.8 &   7.3 & -58.8 & 180.0 & -64.9 & -17.4 & -64.4 &  -1.7  \\
   6.7 &   2.0 &  -1.5 & -64.9 & 405.0 &  89.0 &  -6.4 &   4.5  \\
 -12.1 &  11.5 &  -9.6 & -17.4 &  89.0 & 320.0 &  31.7 &  -9.7  \\
 -10.3 &   4.8 &   4.7 & -64.4 &  -6.4 &  31.7 & 270.0 & -11.4  \\
  37.5 &   7.9 &   1.5 &  -1.7 &   4.5 &  -9.7 & -11.4 & 505.0
\end{array}\right)\textrm{cm}^{-1},
\end{equation}
where the energy of ground state is chosen as $E_0=-12195$
cm$^{-1}$. On account of the manifold of the ground state, the total
Hamiltonian reads $H_{S}^{(1)}=H_{\textrm{SES}}+E_0\vert
G\rangle\langle G\vert$, which governs the quantum dynamics of EET
in the absence of laser fields.


\section*{Acknowledgments}

Qing Ai is supported by National Natural Science Foundation of China
under Grant No.~11505007, the Youth Scholars Program of Beijing
Normal University under Grant No.~2014NT28, and the Open Research
Fund Program of the State Key Laboratory of Low Dimensional Quantum
Physics, Tsinghua University under Grant No.~KF201502. Fu-Guo Deng
is supported by the Fundamental Research Funds for the Central
Universities under Grant No.~2015KJJCA01 and the National Natural
Science Foundation of China under Grant No.~11474026.

\section*{Author contributions statement}

All authors wrote and reviewed the manuscript.
M.J.T. did the calculations. Q.A., Y.C.C. and F.G.D. designed the project.

\section*{Additional information}

Competing financial interests: The authors declare no competing financial interests.

\end{document}